\documentclass[a4paper,11pt]{article}
\usepackage{jinstpub} 
\usepackage{lineno}
\usepackage{subcaption} 

\title{\boldmath Development of a comprehensive PMT optical model for the JUNO experiment}

\author[a,b]{Y. Ren}
\author[a]{X. Yang}
\author[c]{Y. Wang}
\author[a,1]{Z. Deng\note{dengzy@ihep.ac.cn, corresponding author.}}
\author[a]{Z. Qin}
\author[d]{A. Olshevskiy}
\author[e,f]{W. Wang}
\author[d]{N. Anfimov}
\author[a]{Z. Wang}
\author[a,b,2]{G. Cao\note{caogf@ihep.ac.cn, corresponding author.}}

\affiliation[a]{Institute of High Energy Physics, Chinese Academy of Sciences, Beijing 100049, China}
\affiliation[b]{University of Chinese Academy of Sciences, Beijing 100049, China}
\affiliation[c]{Shandong University, Jinan 250100, China}
\affiliation[d]{Joint Institute for Nuclear Research, Dubna, Moscow Region 141980, Russia}
\affiliation[e]{Sino-French Institute of Nuclear Engineering and Technology, Sun Yat-sen University, Zhuhai 519082, China}
\affiliation[f]{School of Physics, Sun Yat-sen University, Guangzhou 510275, China}

\abstract{
There are 17,612 20-inch photomultiplier tubes (PMTs) installed at the Jiangmen Underground Neutrino Observatory (JUNO). Developing a precise optical model for the PMTs is crucial for enhancing the accuracy of detector simulations and refining the energy response model at JUNO. In this study, we established a comprehensive PMT optical model based on the prior studies, taking into account the non-uniformity of the photon detection efficiency (PDE) across the PMT surface and the variances in the PDE as well as reflectance among the different PMTs. By collecting the reflectance data from 669 PMTs and utilizing the PDE data from the mass testing systems, we estimated the thickness maps of the photocathode (PC) and the anti-reflective coating (ARC) for each PMT. We also determined the collection efficiency (CE) by decomposing the PDE with consideration of the optical processes occurring within the PMTs. The refractive index and extinction coefficient of both the PC and ARC, along with the escape factor, were evaluated over a broad wavelength range of 300~nm to 700~nm, covering the entire spectrum of interest for JUNO. Compared to the prediction from a simplified PMT optical model, which assumes the uniform PC and ARC across all the PMTs of the same type, the further developed PMT optical model yields much more reflectance for the HPK PMTs and less for the NNVT PMTs, and the change in the PDE is at the level of a few percent. This comprehensive PMT optical model also provides a valuable reference for other PMT-based applications.
}


\arxivnumber{2601.19081} 

\begin{document}
\maketitle
\flushbottom

\section{Introduction}
\label{sec:intro}

The photomultiplier tube (PMT) has served as an optimal photosensor for numerous neutrino projects over the past few decades and remains pivotal in current and future initiatives. One such project is the Jiangmen Underground Neutrino Observatory (JUNO), which has deployed 17,612 20-inch PMTs and 25,600 3-inch PMTs in its central detector. The spherical arrangement of these PMTs achieves up to 75\% photocathode (PC) coverage. Among the 20-inch PMTs, about 5,000 are dynode PMTs manufactured by Hamamatsu Photonics K.K. (HPK), while the remainder are MCP-PMTs produced by Northern Night Vision Technology Co. (NNVT) \cite{WANG2012113}. These PMTs are responsible for capturing light from the 20 kton liquid scintillator contained within an acrylic sphere with a diameter of 35.4~m. The detailed design of the JUNO detector is described in~\cite{JUNO:2015sjr}. JUNO's primary goal is to determine the neutrino mass ordering and precisely measure the neutrino oscillation parameters by analyzing oscillation patterns within the neutrino energy spectrum from the Taishan and Yangjiang nuclear power plants, located approximately 53~km from the JUNO site. Additionally, JUNO has an extensive physics program focused on detecting neutrinos from various natural sources, as outlined in~\cite{JUNO:2015zny,JUNO:2022hxd}.

A precise understanding of the detector response is crucial for determining the neutrino mass ordering and conducting other physics analyses in JUNO~\cite{JUNO:2024jaw}. Specifically, the detector response primarily involves the energy resolution, energy non-uniformity, and energy non-linearity within the JUNO detector. The responses of PMTs, particularly the photon detection efficiency (PDE), significantly influence these energy responses and must be carefully managed. Understanding the PDE of PMTs is complex, as it depends on the various factors, including the working media, the angle of incidence (AOI) of photons, photon energy, positions where the light hits the PC, and the shape as well as the internal structure of PMTs. The influence of these factors on the PDE is discussed in detail in~\cite{Wang:2022tij}. Within the JUNO experiment, the PDE of 20-inch PMTs has been thoroughly characterized via the PMT mass testing system, including the average PDE measured by the container system for all the PMTs and the PDE uniformity measured by the scanning station system for a subset, with results summarized in~\cite{JUNO:2022hlz}. However, the measured PDE cannot be directly utilized in detector simulations to predict energy responses because the PMTs operate in water within the JUNO detector, rather than in air, and the AOIs, energies, and positions of photons hitting the PMTs differ from those in the mass testing system. In fact, these factors can vary significantly from event to event in the JUNO detector and lead to the different PDE. Therefore, a PMT optical model is essential to accommodate all these factors while adhering to constraints from the mass testing data.

A new optical model for the PMTs has been proposed in~\cite{Wang:2022tij}, which accounts for the PMT window as a composite structure comprising the glass substrate, a thin anti-reflective coating (ARC), and a PC. The key parameters of this model were determined from detailed measurements performed on three 20-inch PMTs: two from NNVT and one from HPK. By incorporating the optical properties of the PC and ARC extracted from these measurements into the JUNO simulation program, the energy response of the JUNO detector has been evaluated and predicted under the assumption that PMTs of the same type with the identical optical properties of the PC and ARC~\cite{JUNO:2024fdc}. However, the variations in the PDE exist among PMTs, and the properties of the PC and ARC could be non-uniform. Moreover, the PMT reflectance exhibits the significant variations, which must also be taken into account, as light reflected from the PMTs has a substantial impact on the light yield and energy response in the JUNO detector. In this study, these effects are incorporated by extending the PMT optical model originally developed in~\cite{Wang:2022tij}. The extension utilizes the PDE data acquired from both the container system and the scanning station system, as well as the reflectance data collected in this study. The resulting comprehensive PMT optical model enables the decoupling of the quantum efficiency (QE) and collection efficiency (CE) from the measured PDE, while accounting for the non-uniformity of the PC and ARC. Specifically, the QE is determined based on the photon absorption ratio in the PC and the escape factor, which represents the fraction of photoelectrons that successfully escape from the PC. Utilizing this model, we derived the thickness maps of the PC and ARC for all 17,612 20-inch PMTs, along with their corresponding CE values. The model covers the entire range of the photon wavelengths relevant to JUNO, from 300~nm to 700~nm. This comprehensive approach also provides valuable insights for other PMT-based projects seeking to improve their detector performance.

The subsequent sections are organized as follows: we first review the PMT optical model developed in~\cite{Wang:2022tij}, followed by a detailed discussion of the comprehensive PMT optical model. We then introduce the PDE data utilized in this study, and illustrate the method employed for the PMT reflectance measurements and report the reflectance results. Subsequently, the extracted basic parameters for the PMT optical simulation are discussed, including the thickness maps of the PC and ARC and CE for individual PMTs, as well as the refractive index and extinction coefficient over an extended wavelength range for three types of PMTs. In the end, the comparison of the average reflectance and PDE calculated from the simplified and comprehensive models is presented.

\section{A comprehensive PMT optical model}
\label{sec:pmt_model}
\subsection{Recap of the PMT optical model}
\label{sec:model_recap}
The PDE could be denoted as a function of the photon wavelength ($\lambda$), the AOI ($\alpha$), and the position ($\vec{r}$) of the photon hitting the PC of PMTs. Specifically, the measured PDE can be contributed by the multiple positions in addition to the point of incidence on the PC, as a fraction of the incident light could penetrate the PC and propagate within the PMTs. These transmitted photons could hit the PC again and then be absorbed, or even undergo multiple reflections before absorption. Consequently, the PDE can be expressed as shown in Eq.~\ref{eq:DE_Model} and is discussed in detail in~\cite{Wang:2022tij}:
\begin{equation}
        \mathrm{PDE}(\lambda, \alpha, \vec{r})=\sum_{j}a_j(\lambda, \alpha, \vec{r})\cdot\beta_j\cdot\rho_j(\lambda, \vec{r})\cdot \mathrm{CE}_j(\vec{r}) = \sum_{j}a_j(\lambda,\alpha, \vec{r})\cdot\beta_j\cdot F_j(\lambda, \vec{r})
        \label{eq:DE_Model}
\end{equation}
Here, $j$ represents the $j$-th photon, and $a$ represents the absorption ratio of the incident light, and $\rho$ is the escape factor. The product of $\rho$ and CE is denoted as an $F$-factor. The coefficient $\beta$ weights the absorption from the different positions that contribute to the measured PDE. For the ideally uniform PC and ARC, the positional dependencies of $a$ and $\rho$ would be negligible. While this assumption may hold in certain applications, it is generally not valid for large PMTs, such as those employed in JUNO. The realistic non-uniform optical parameters of the PC and ARC could lead to the non-uniformity of the absorptance, which contributes to the non-uniform PDE.

The CE is determined by the electric field within the PMTs and represents the probability of detecting a photon, which typically varies with the positions on the PC, contributing to the PDE non-uniformity as well. By selecting several reference positions with known \(F\)-factors on the PC, the \(F\)-factor at any other position can be derived through interpolation. Therefore, Eq.~\ref{eq:DE_Model} can be rewritten as:
\begin{equation}
        \mathrm{PDE}=\sum_{j}a_j \sum_{k=1}^{n}\beta_{jk} F_k
        \label{eq:DE_Model_final}
\end{equation}
where \(\beta\) represents the weight coefficient, and \(F_k\) denotes the \(F\)-factor at the \(k\)-th reference point. A procedure for deriving the \(F\)-factors at these reference points has been proposed in~\cite{Wang:2022tij}. The coefficient \(\beta\) can be computed by simulating the optical processes inside the PMTs utilizing a simulation toolkit. The absorptance, reflectance, and transmittance of the incident photons on the PC can be calculated based on the optical theory with the optical properties of the PC and ARC, including the refractive index (\(n\)), extinction coefficient (\(k\)), and thickness (\(d\)). In the previous work, the reflectance data collected by immersing PMTs in the linear alkylbenzene are utilized to extract values for \(n\), \(k\), and \(d\) for two NNVT PMTs and one HPK PMT across a wavelength range of 390 nm to 500 nm. These PMTs represent three different types of 20-inch PMTs equipped in the JUNO detector. For the NNVT PMTs, there are two types known as NNVT normal-QE PMT and NNVT high-QE PMT, which were produced during the early stages of the PMT mass production and manufactured with the improved QE achieved through the optimized PC technology, respectively.

\subsection{Extension of the PMT optical model}
\label{sec:model_extend}
The following assumptions have been made to enable the PMT optical model to account for the PDE non-uniformity within a single PMT and the PDE variations among different PMTs:
\begin{itemize}
\item The PMTs of the same type share the identical spectral responses for the refractive index and extinction coefficient of both the PC and ARC. The values of $n$ and $k$, as well as the escape factor, acquired in~\cite{Wang:2022tij},  are employed in this study for the three types of PMTs, covering the wavelengths from 390~nm to 500~nm. Utilizing dispersion relations and available data in the literature, we extend this range down to 300~nm and up to 700~nm, encompassing the relevant spectrum for JUNO.
\item The variations in the QE and reflectance for a single PMT are induced by the non-uniform thicknesses of the PC and ARC. The thickness map is determined in this work according to the optical theory, with inputs derived from the PDE and reflectance data for each 20-inch PMT. These inputs will be discussed in detail in Sec.~\ref{sec:pmt_data}.
\item Due to the absence of the specific QE information, decoupling the position dependence of the QE and CE from the measured PDE exists some problems. Therefore, we assume that the CE remains constant in the PC region for the zenith angles smaller than 70 degrees (in regions far from the equator) when calculating the thickness maps of the PC and ARC. Thereafter, the positional dependence of the CE and its variations among different PMTs are incorporated into Eq.~\ref{eq:DE_Model_final}.
\item For simplification, we assume that both the QE and CE are independent of the azimuthal angles on the PC. This assumption is a reasonable approximation for the QE of both the NNVT PMTs and HPK PMTs, as demonstrated in~\cite{JUNO:2022hlz}. Regarding the CE, while it is azimuthal angle-dependent for the HPK PMTs near the equator due to the orientation of the focusing electrode, the random arrangement of these electrodes during the PMT installation in the JUNO detector largely mitigates their impact on energy responses, as the contributions from multiple PMTs during physical events sufficiently average out these variations.
\end{itemize}

\begin{figure}[htbp]
\centering
\includegraphics[width=0.9\textwidth]{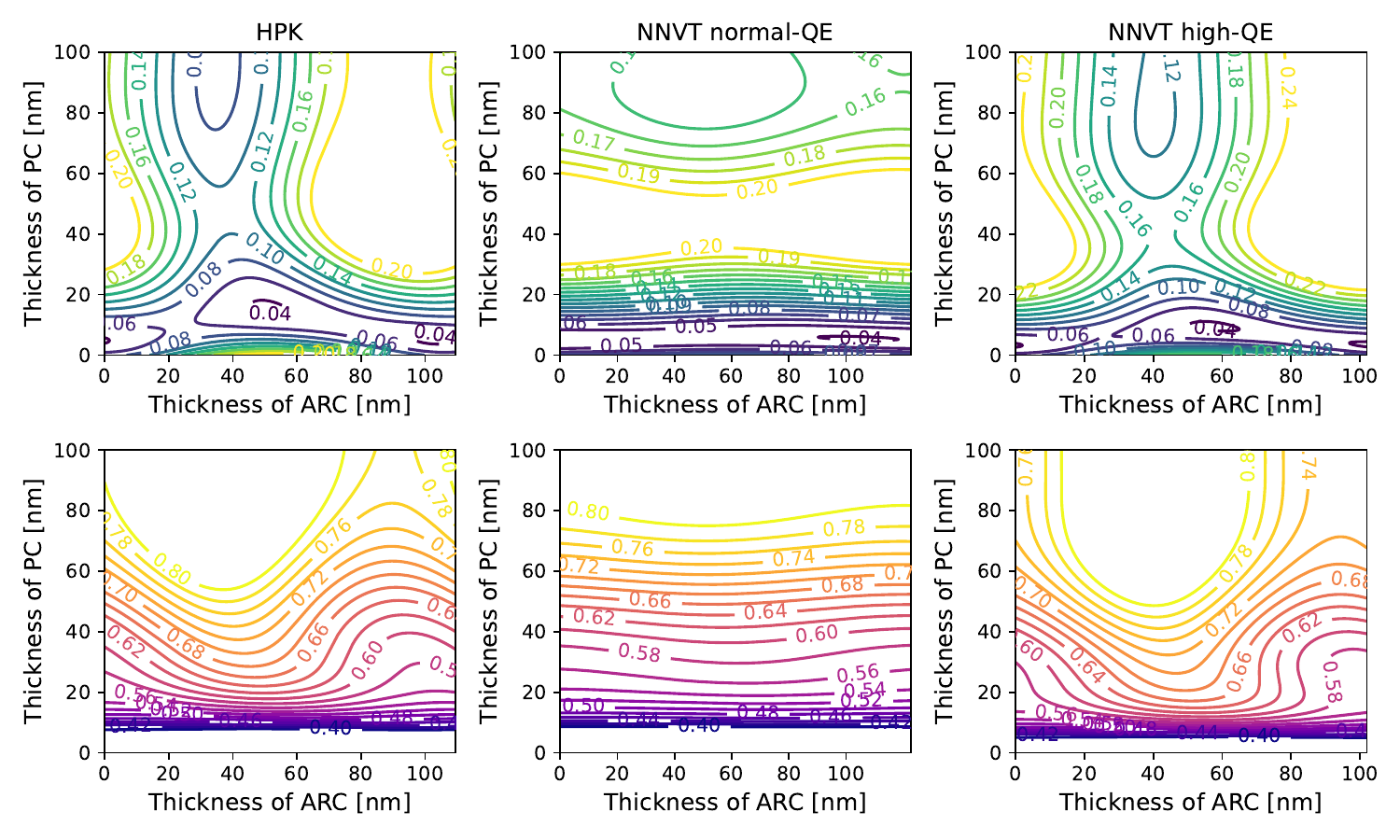}
\caption{Contours of the reflectance (top) and absorptance (bottom) at 415~nm and 420~nm, respectively, as a function of the thicknesses of the ARC (x-axis) and PC (y-axis) for the HPK PMT (left), NNVT normal-QE PMT (middle) and NNVT high-QE PMT (right).}
\label{fig:contour_r_a}
\end{figure}

Given the assumptions above, the thickness map of the PC and ARC can be derived by simultaneously fitting the PDE and reflectance data for all the 20-inch PMTs. To model the position dependence of the thickness for both the PC and ARC, an empirical formula is employed:
\begin{equation}
        d_{\mathrm{PC},\mathrm{ARC}} = c_{2_{\mathrm{PC},\mathrm{ARC}}} \times \theta^2 + c_{1_{\mathrm{PC},\mathrm{ARC}}} \times \theta + c_{0_{\mathrm{PC},\mathrm{ARC}}}
        \label{eq:d_dep}
\end{equation}
where \(\theta\) represents the zenith angle of the position on the PC, while \(c_0\), \(c_1\), and \(c_2\) are free parameters determined through the fitting process. Fig.~\ref{fig:contour_r_a} illustrates the contours of the reflectance (top) and absorptance (bottom) at 415~nm and 420~nm, respectively, calculated as a function of the thicknesses of the ARC (x-axis) and PC (y-axis) for the HPK PMT (left), NNVT normal-QE PMT (middle), and NNVT high-QE PMT (right). The refractive index and extinction coefficient of the ARC and PC reported in~\cite{Wang:2022tij} are used in these calculations. Notably, the different types of PMTs exhibit distinct dependencies in their thickness maps. Once the thickness maps of the PC and ARC are established, the non-uniformity of the absorptance on the PC can be calculated. The weight coefficients are computed by simulating the optical processes within the PMTs utilizing a simulation program developed based on Geant4. Subsequently, the CE at the reference points can be determined with the PDE data at those points via Eq.~\ref{eq:DE_Model_final}, and then the CE at other points can be derived by interpolation.

\section{PDE and reflectance data}
\label{sec:pmt_data}
In this study, both the PDE results measured by the mass testing systems including the container system and the scanning station system and the newly collected reflectance data are utilized to constrain the thicknesses of the PC and ARC. 

\subsection{PDE data}
For each 20-inch PMT, the PDE is measured with a container system that employs a 420~nm light source to illuminate the entire PMT PC in air. Additionally, a subset of PMTs is evaluated for the PDE uniformity test with a scanning station system, which scans the PC with the seven LEDs that provide 420~nm light nearly perpendicular to the surface at the seven fixed zenith angles, meanwhile the LEDs can be rotated to cover the different azimuth angles. The PDE obtained from the container system is then corrected to align with the PDE as defined by the scanning station system, which averages the PDE values acquired at the positions of the seven LEDs, weighted by their respective surface areas, and Table~\ref{tab:led_weights} summarizes the surface area weights along with the zenith angles for each LED for both the NNVT and HPK PMTs. In this study, the PDE values averaged over the azimuthal angles for the seven zenith angles are utilized to extract the thickness maps of the PC and ARC. 

\begin{table}[htbp]
\centering
\caption{The zenith angles and the PC area weights of the seven LEDs equipped in the scanning station system~\cite{JUNO:2022hlz}.}
\label{tab:led_weights}
\smallskip
\begin{tabular}{|c|c|c|c|c|c|c|c|}
\hline
 & LED1 & LED2 & LED3 & LED4 & LED5 & LED6 & LED7\\
\hline
Zenith angle (NNVT) & 14$^\circ$ & 30$^\circ$ & 42.5$^\circ$ & 55$^\circ$ & 67$^\circ$ & 77.5$^\circ$ & 85$^\circ$ \\
\hline
Area weight (NNVT) & 4.85\% & 8.97\% & 12.58\% & 17.21\% & 19.92\% & 18.04\% & 18.43\%  \\
\hline
Zenith angle (HPK) & 13$^\circ$ & 28$^\circ$ & 41$^\circ$ & 55$^\circ$ & 66$^\circ$ & 79$^\circ$ & 85$^\circ$ \\
\hline
Area weight (HPK) & 4.51\% & 8.76\% & 13.49\% & 17.11\% & 20.46\% & 18.65\% & 17.03\%   \\
\hline
\end{tabular}
\end{table}

However, since only approximately 3,200 PMTs were evaluated with the scanning station system, it is necessary to develop a method to estimate the PDE values at the seven zenith angles for the PMTs that were not tested in this system in order to simulate their PDE non-uniformity as well. A straightforward approach to predict these PDE values is to perform random sampling based on the measured data. To account for the potential dependence of the PDE on zenith angles, a seven-dimensional PDE histogram is constructed according to data collected with the scanning station system for each type of PMT, which includes 585 HPK PMTs, 1,719 NNVT normal-QE PMTs, and 939 NNVT high-QE PMTs. The random sampling of the PDE is carried out with the seven-dimensional PDE histograms, which allows the determination of the PDE values at the seven reference positions for all the PMTs that were not tested with the scanning station system. A normalization factor is applied to the seven sampled values during the sampling process to ensure that the average sampled PDE of each PMT aligns with the PDE results acquired from the container system, where the average PDE is calculated by applying the weights presented in Table~\ref{tab:led_weights}.

\begin{figure}[htbp]
\centering
\includegraphics[width=0.9\textwidth]{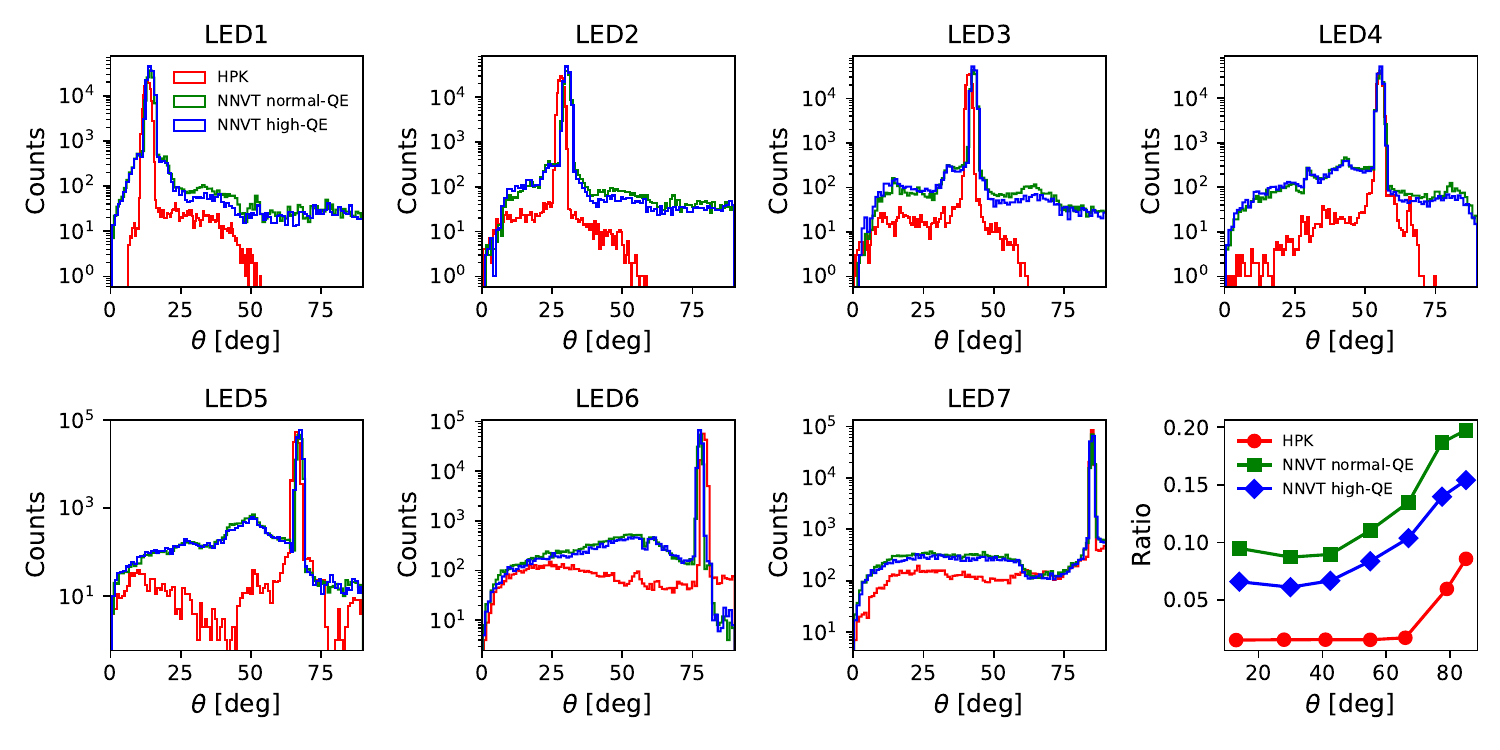}
\caption{The zenith angle distributions of the absorbed photons for the seven LEDs for the HPK PMTs shown in red, NNVT normal-QE PMTs in green, and NNVT high-QE PMTs in blue, and the ratio of the absorbed transmitted photons to the total absorbed photons is shown in the last panel.}
\label{fig:pde_compose}
\end{figure}

Since the PDE obtained above encompasses the contributions from the transmitted photons absorbed at the multiple positions on the PC in addition to the direct absorption at the incident position, a detailed PMT optical simulation is conducted to distinguish the absorbed positions of the photons. Utilizing the PMT optical simulation software developed in~\cite{Wang:2022tij}, the detailed structure of the seven LED light sources has been newly implemented in the software, reflecting their real locations and orientations in the scanning station system, which allows for accurate accounting of the distributions of the AOI and the beam profile of light impinging on the PMT surface. The simulations for the seven LEDs were conducted with the uniform optical properties of the PC and ARC acquired from~\cite{Wang:2022tij}. Fig.~\ref{fig:pde_compose} illustrates the zenith angle distributions of the absorbed photons for the seven LEDs for the HPK PMTs shown in red, NNVT normal-QE PMTs in green, and NNVT high-QE PMTs in blue. This figure indicates that the total absorption primarily arises from the incident positions of the light beams, which are represented by the peak, while the rest shows the contributions from the transmitted photons. The ratio of the absorbed transmitted photons to the total absorbed photons is depicted in the last panel and applied to correct the PDE by excluding these extra contributions based on their weights. It is observed that the two types of the NNVT PMTs exhibit the similar distributions due to their same shape and structure, and compared to the HPK PMTs, they gain significantly larger contributions from the transmitted photons.

\subsection{Reflectance data}
The reflectance measurements were carried out, utilizing the setup developed in this work. In the first stage, a total of 565 20-inch PMTs were randomly selected for testing, which included 276 HPK PMTs, 115 NNVT normal-QE PMTs, and 174 NNVT high-QE PMTs. The reflectance was measured at 415 nm across the four different zenith angles, including 0$^\circ$, 28$^\circ$, 51$^\circ$ and 72$^\circ$. In the second stage, only the PMTs tested with the scanning station system were selected, and measured at the four zenith angles as well as the azimuth angles scanned with a step of 45$^\circ$, involving 21 HPK PMTs, 38 NNVT normal-QE PMTs, and 45 NNVT high-QE PMTs.

\begin{figure}[htbp]
\centering
\includegraphics[width=0.8\textwidth]{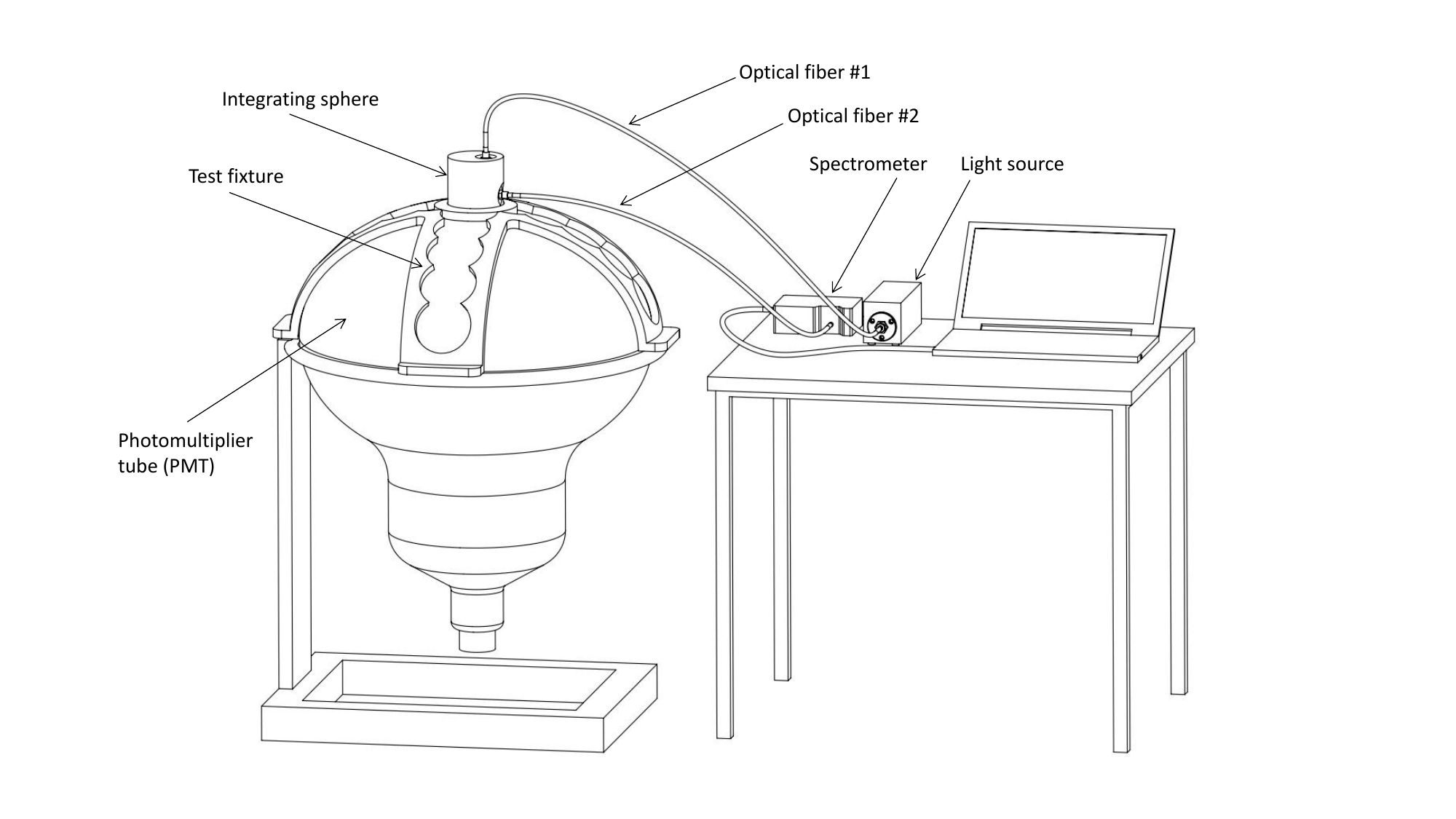}
\caption{The schematic diagram of the reflectance setup.}
\label{fig:ref_setup}
\end{figure}

\begin{figure}[htbp]
\centering
\includegraphics[width=0.8\textwidth]{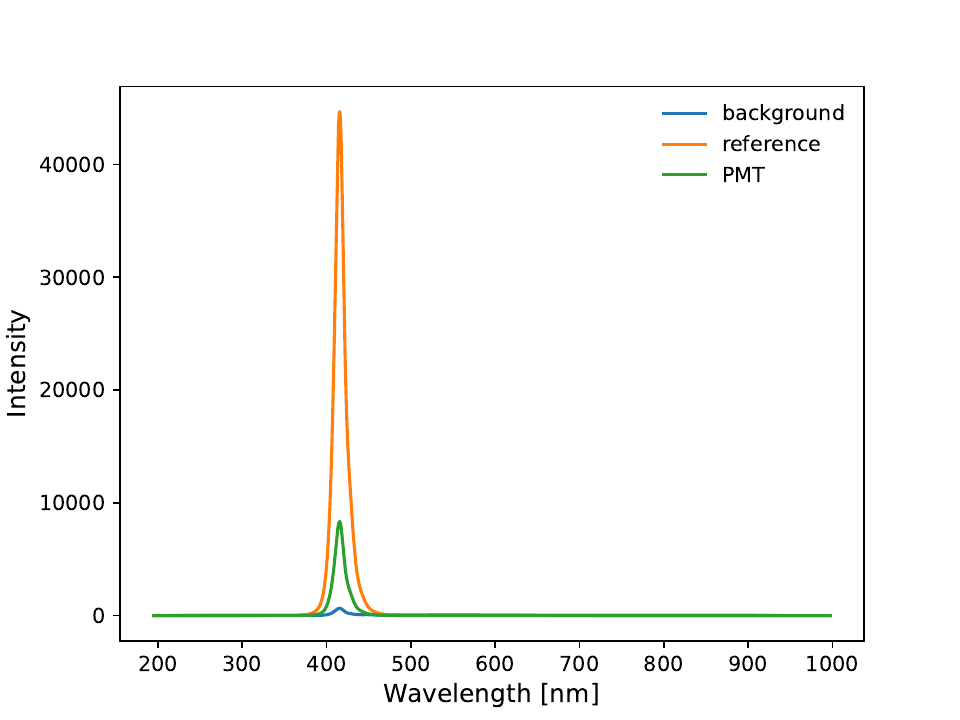}
\caption{An example of the measured spectra: the background (blue), the PMT (green), and the reference sample (orange).}
\label{fig:ref_spec_example}
\end{figure}

The schematic diagram of the reflectance measurement setup is shown in Fig.~\ref{fig:ref_setup}. The system consists of an LED light source coupled with an optical fiber, with a central wavelength of 415 nm and a full width at half maximum of 14 nm. An integrating sphere, the ISP-50-8-REFL from Mikropack, is employed to collect the reflected light from the tested samples, along with a spectrometer and a computer for data storage, while a testing mold provides the positional information on the PMTs. In the setup, optical fiber \#1 transmits light from the LED to the integrating sphere, where the light beam is collimated by a lens integrated at the top of the sphere. The incident light beam exits the sphere at an AOI of 8$^\circ$ through a hole at the bottom, which has a diameter of approximately 8 mm. When a sample is placed over this hole, the reflected light is collected by the integrating sphere, and a fraction of this light is transmitted through optical fiber \#2 to the spectrometer, where its intensity is measured as a function of the wavelength. By comparing the measured intensity of the test sample with the reference sample of the known reflectance, the reflectance of the test sample can be determined. This measurement does not require dark conditions, as the background spectrum can be directly measured without samples placed over the hole of the integrating sphere. Fig.~\ref{fig:ref_spec_example} illustrates an example of the measured spectra: the background (blue), the PMT (green), and the reference sample (orange).

The systematic uncertainty has been carefully estimated, with the relative uncertainty determined to be approximately 2.5\%. The primary contributions to this uncertainty arise from the LED stability and the reflectance uncertainty of the reference sample. Additionally, the reflectance of PMT was corrected by the ratio of the reflectance of the flat sample to the curved one for each zenith angle to eliminate the influence of the curvature of the PMT surface on the reflectance. Since no dependence of the reflectance on the azimuthal direction was observed, the reflectance values measured at the different azimuth angles are finally averaged for each zenith angle.

\begin{figure}[htbp]
\centering
\includegraphics[width=0.9\textwidth]{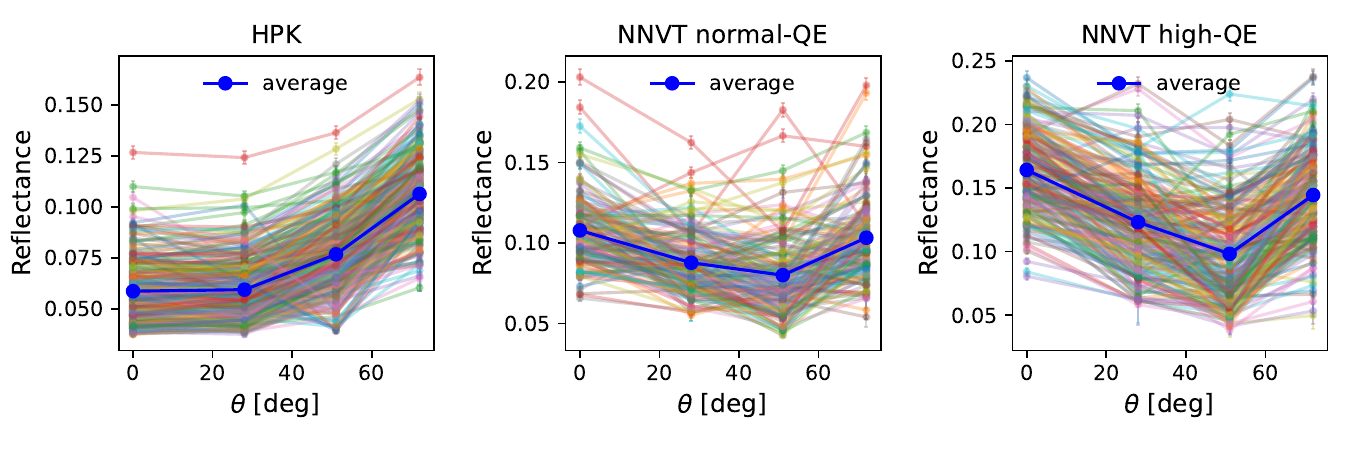}
\caption{The measured reflectance of the HPK PMTs (left), NNVT normal-QE PMTs (middle), and NNVT high-QE PMTs (right). The solid blue lines shown in the figure represent the average reflectance for each type of PMT.}
\label{fig:measured_ref}
\end{figure}

The measured results are presented in Fig.~\ref{fig:measured_ref}, showing data for all the tested HPK PMTs (left), NNVT normal-QE PMTs (middle), and NNVT high-QE PMTs (right), and the solid blue lines represent the average reflectance for each type of PMT. The significant variations in the reflectance can be observed across all the types of PMTs. Specifically, the average reflectance at the zenith angle of zero degree for the HPK PMTs is approximately 5.9\%, which is lower than 10.8\% and 16.4\% observed for the NNVT normal-QE and NNVT high-QE PMTs, respectively. Additionally, the reflectance of the HPK PMTs increases in areas closer to the equator; in contrast, the reflectance of the NNVT PMTs exhibits a decreasing trend from 0$^\circ$ to about 50$^\circ$, followed by an increase. 

A similar random sampling approach to that applied for estimating the PDE is employed to assess the reflectance of PMTs that have not been measured. A four-dimensional reflectance histogram is constructed with the measured data for each type of PMT to account for the positional dependence of the reflectance. Since the collected data indicate a weak correlation between the reflectance and PDE, this relationship is not considered in the aforementioned sampling process.

\section{Results}
\subsection{Thickness map of the PC and ARC}
A chi-squared function is constructed to simultaneously fit the reflectance and PDE data:
\begin{equation}
\label{eq:chi2}
\begin{aligned}
        \chi^2 = \sum_{i=1}^4 \left( \frac{R(\theta_i) - R_{theo}(d_{\mathrm{PC}}(\theta_i), d_{\mathrm{ARC}}(\theta_i))}{\sigma_R} \right)^2 + \\
        \sum_{j=1}^5 \left( \frac{A(\theta_j) - A_{theo}(d_{\mathrm{PC}}(\theta_j), d_{\mathrm{ARC}}(\theta_j))}{\sigma_A} \right)^2 + \\
        \left( \frac{\mathrm{CE'} - \mathrm{CE}_{theo}}{\sigma_{\mathrm{CE'}}} \right)^2
\end{aligned}
\end{equation}
In this equation, \(R(\theta_i)\) represents the measured reflectance at the four zenith angles, while \(R_{theo}\) and \(A_{theo}\) are the calculated reflectance and absorptance, respectively, based on the optical theory. The predicted \(\mathrm{CE}_{theo}\) is set to 0.93 for the HPK PMTs and 1.0 for the NNVT PMTs. The uncertainties \(\sigma_R\), \(\sigma_A\), and \(\sigma_{\mathrm{CE'}}\) correspond to the measured reflectance, expected absorptance, and assumed constant CE', respectively. Here, \(\sigma_{\mathrm{CE'}}\) is assumed to be 3\%, and \(\sigma_A\) is estimated through error propagation from the measured PDE and its uncertainty, as the absorptance \(A(\theta_j)\) can be calculated via:
\begin{equation}
\label{eq:absorb}
    A(\theta_j) = \frac{\mathrm{PDE}'(\theta_j)}{\rho \times \mathrm{CE'}}
\end{equation}
where \(\mathrm{PDE}'\) represents the corrected PDE, which excludes the contributions from the absorbed transmitted photons to the total PDE. Only the PDE measured at the five small zenith angles, namely from LED1 to LED5, are utilized in the fitting process. The value of $\rho$ is at 420~nm from~\cite{Wang:2022tij}.

\begin{figure}[htbp]
\centering
\includegraphics[width=0.9\textwidth]{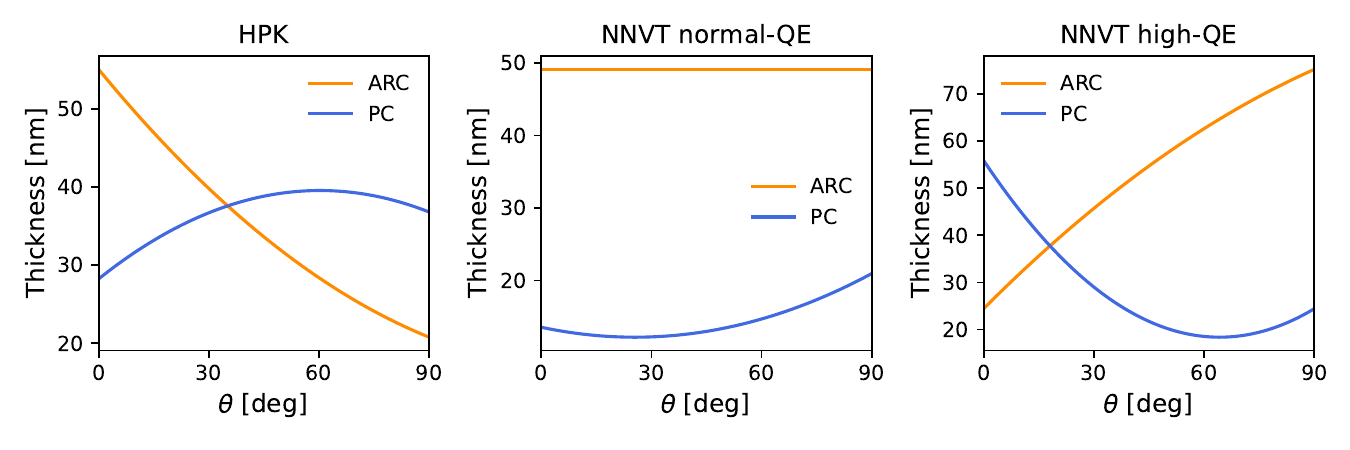}
\caption{A typical example of the thickness curve of the PC and ARC as a function of the zenith angles for the HPK PMTs (left), NNVT normal-QE PMTs (middle), and NNVT high-QE PMTs (right).}
\label{fig:thickness_curve}
\end{figure}

\begin{figure}[htbp]
\centering
\includegraphics[width=0.9\textwidth]{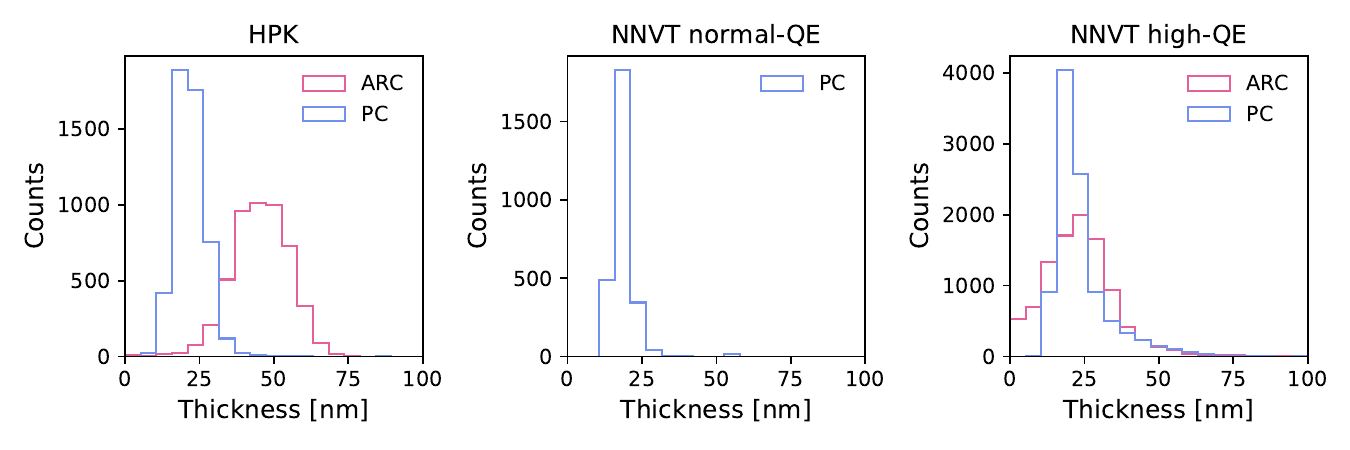}
\caption{Thickness distributions of the PC and ARC at the zenith angle of zero degree for all the HPK PMTs (left), NNVT normal-QE PMTs (middle), and NNVT high-QE PMTs (right).}
\label{fig:thickness_distr}
\end{figure}

By minimizing Eq.~\ref{eq:chi2}, the free parameters \(c_0\), \(c_1\), and \(c_2\) in Eq.~\ref{eq:d_dep} can be constrained for both the PC and ARC of each PMT. An example of the thickness curves derived from the fitting is illustrated in Fig.~\ref{fig:thickness_curve} for the HPK PMT (left), NNVT normal-QE PMT (middle), and NNVT high-QE PMT (right). The thickness of the ARC has been fixed to 49.1 nm for the NNVT normal-QE PMTs, due to its negligible effect on the reflectance and absorptance, as demonstrated in Fig.~\ref{fig:contour_r_a}. The obtained thickness distributions at the zenith angle of zero are presented in Fig.~\ref{fig:thickness_distr} for all the HPK PMTs (left), NNVT normal-QE PMTs (middle), and NNVT high-QE PMTs (right).

\subsection{CE}
The aforementioned PMT optical simulation is conducted again for the seven LEDs in the scanning station system by incorporating the constructed thickness map for each PMT into the simulation. This allows us to determine the contributions from the various positions on the PC to the PDE, taking into account the PC and ARC non-uniformity. From the simulation, both \(a\) and \(\beta\) can be derived for the seven reference positions. Applying Eq.~\ref{eq:DE_Model_final}, the CE at these reference positions can be calculated for each PMT by solving the system of equations with the PDE data at each point. 

\begin{figure}[htbp]
\centering
\includegraphics[width=0.9\textwidth]{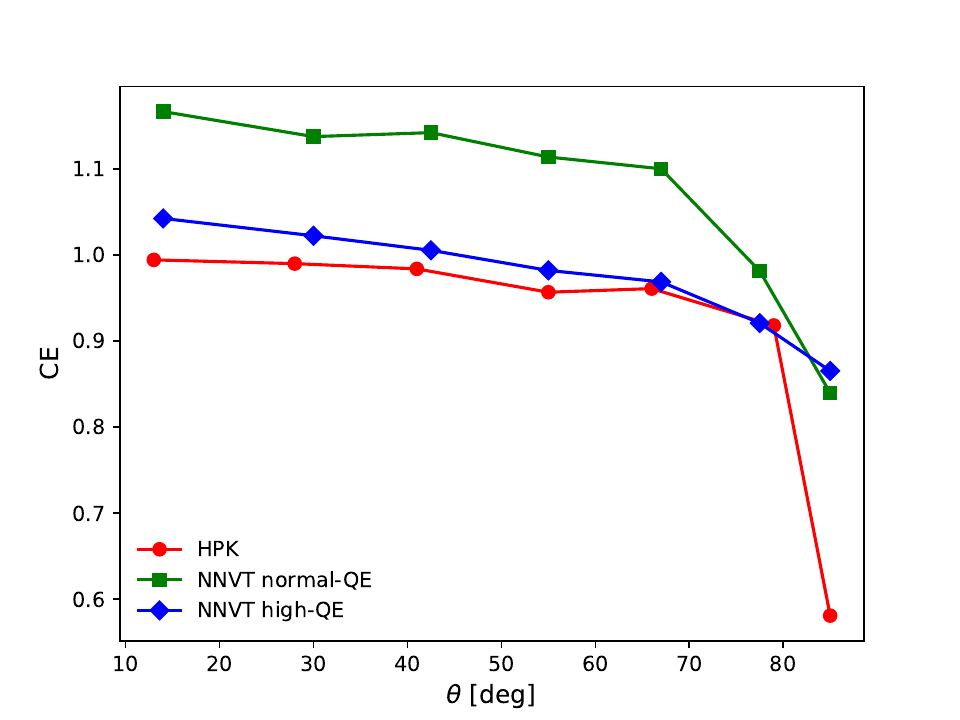}
\caption{The average CE as a function of the zenith angle for the HPK PMTs (red), NNVT normal-QE PMTs (green), and NNVT high-QE PMTs (blue).}
\label{fig:average_ce}
\end{figure}

Fig.~\ref{fig:average_ce} presents the average CE across all the PMTs of the same type, illustrating the CE decreases with the increasing zenith angle. It is noted that some CE values are larger than one, which is a consequence of ignoring the variations of the escape factor among PMTs with the assumption that the escape factor spectra remains unchanged for each type of PMT, since decoupling the CE from the escape factor based on the limited data exits challenges in this work.

\subsection{Spectral response of optical properties}
\begin{figure}[htbp]
\centering
\begin{subfigure}[b]{0.3\textwidth}
\includegraphics[width=\textwidth]{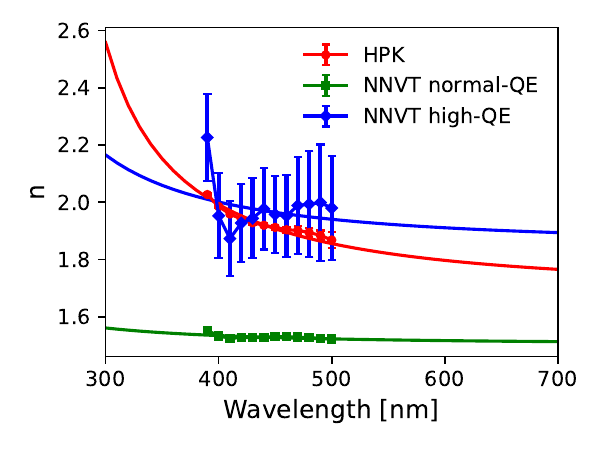}
\caption{}
\label{fig:optical_properties_a}
\end{subfigure}
\hfill
\begin{subfigure}[b]{0.3\textwidth}
\includegraphics[width=\textwidth]{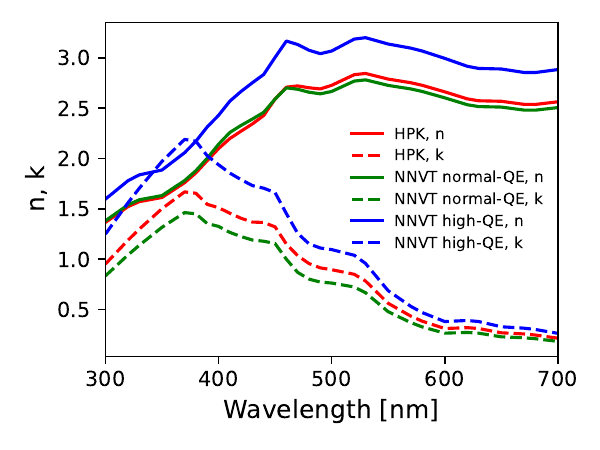}
\caption{}
\label{fig:optical_properties_b}
\end{subfigure}
\hfill
\begin{subfigure}[b]{0.3\textwidth}
\includegraphics[width=\textwidth]{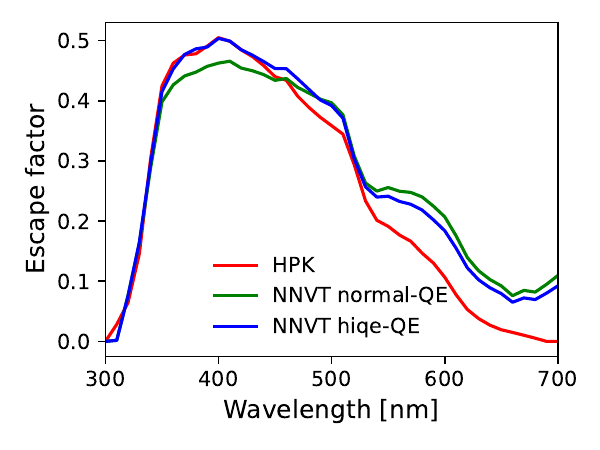}
\caption{}
\label{fig:optical_properties_c}
\end{subfigure}
\caption{(a) The refractive index of the ARC as a function of the wavelength for the HPK PMT (red), NNVT normal-QE PMT (green), and NNVT high-QE PMT (blue). (b) The refractive index (solid lines) and extinction coefficient (dashed lines) of the PC as a function of the wavelength for the HPK PMTs (red), NNVT normal-QE PMTs (green), and NNVT high-QE PMTs (blue). (c) The escape factor as a function of the wavelength for the HPK PMT (red), NNVT normal-QE PMT (green), and NNVT high-QE PMT (blue).}
\label{fig:optical_properties}
\end{figure}

The refractive index and extinction coefficient were measured for the PC and ARC of each type of PMT in~\cite{Wang:2022tij} within the wavelength range of 390~nm to 500~nm. In this study, the dispersion relation is utilized to extend the refractive index of the ARC, covering the wavelength range of interest for JUNO from 300~nm to 700~nm:
\begin{equation}
    n^2 - 1 = \frac{p_0 \lambda^2}{\lambda^2 - p_1}
    \label{eq:dispersion}
\end{equation}
where \(p_0\) and \(p_1\) are determined by fitting the measured refractive index of the ARC between 390~nm and 500~nm. The results of the refractive index of the ARC for the three types of PMTs are shown in Fig.~\ref{fig:optical_properties_a}.

For the PC, the refractive index and extinction coefficient in the wavelength ranges of 300~nm to 390~nm and 500~nm to 700~nm are calculated with the permittivity ($\epsilon$) of the bialkali PC reported in~\cite{HARMER2006439} with the following formula:
\begin{equation}
    \epsilon = \epsilon' + i\epsilon'' = n^2 - k^2 + i2nk
    \label{eq:permitt}
\end{equation}
The calculated values in these two ranges are then normalized to the measured values at 390~nm and 500~nm as found in~\cite{Wang:2022tij}. Fig.~\ref{fig:optical_properties_b} illustrates the results for refractive index (solid lines) and extinction coefficient (dashed lines) across the full wavelength range for the HPK PMT (red), NNVT normal-QE PMT (green), and NNVT high-QE PMT (blue).

The escape factor in the ranges of 300~nm to 390~nm and 500~nm to 700~nm is computed based on the measured QE spectra in~\cite{JUNO:2024fdc} and the respective absorption spectra, which is calculated with the aforementioned extended refractive index and extinction coefficient as well as the thickness values in~\cite{Wang:2022tij} for both the PC and ARC, and then the escape factors are normalized to align with the measured curve reported in~\cite{Wang:2022tij}. The results for the escape factor are shown in Fig.~\ref{fig:optical_properties_c} for the three types of PMTs.

\subsection{Comparison of the average reflectance and PDE}
To evaluate the potential impacts of the comprehensive PMT optical model on the JUNO detector’s energy response, we compare the average reflectance and PDE derived from the comprehensive model to those obtained utilizing the simplified PMT optical model, where all the PMTs of the same type are assumed to share identical optical properties for the PC and ARC as described in~\cite{Wang:2022tij}. 

\begin{figure}[htbp]
\centering
\includegraphics[width=0.9\textwidth]{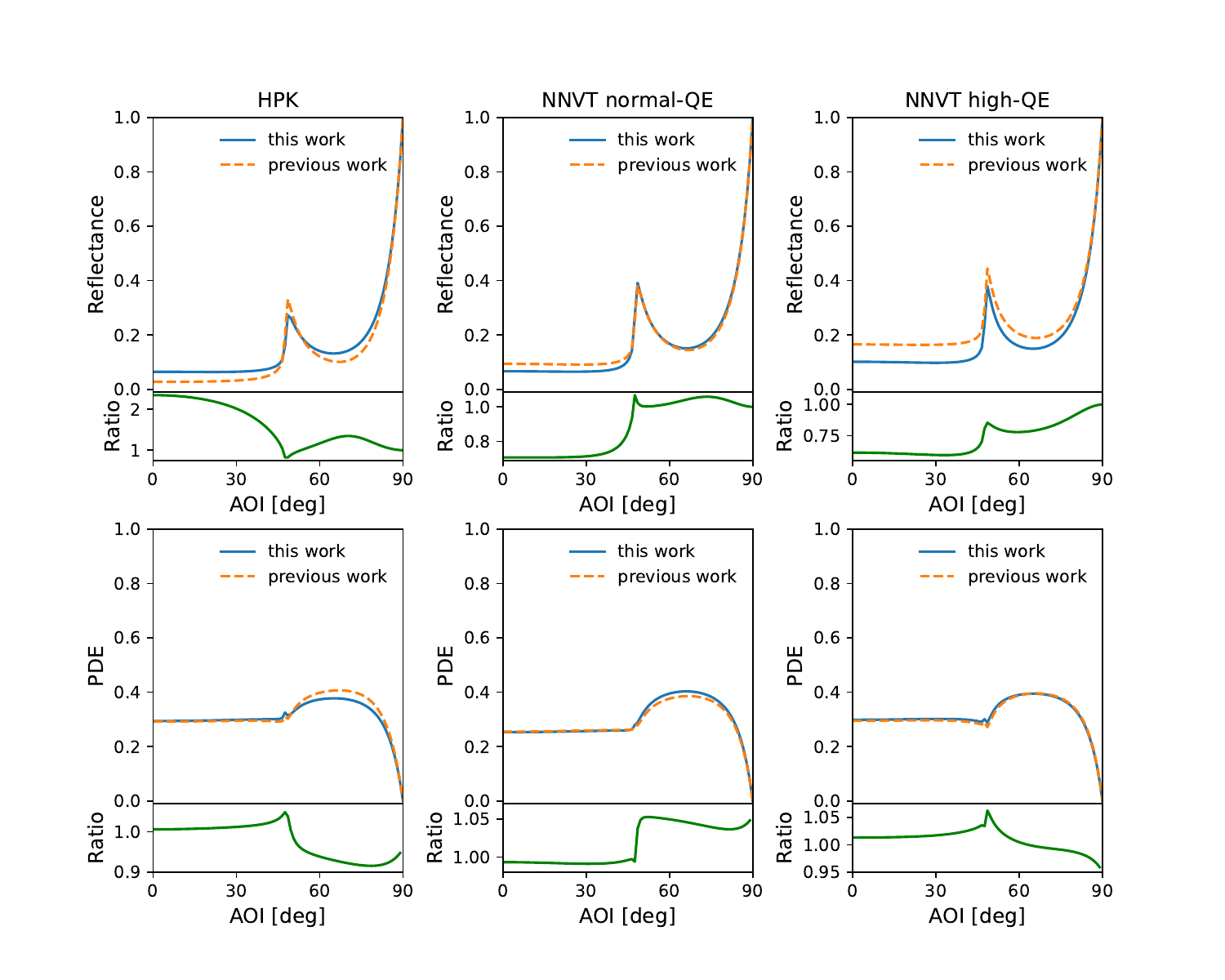}
\caption{The average reflectance (top) and PDE without the contributions from the transmitted photons (bottom) as a function of the AOI for the HPK PMT (left), NNVT normal-QE PMT (middle) and NNVT high-QE PMT (right). The blue solid lines represent the results from the non-uniform ARC and PC, while the orange dashed lines indicate the results from the uniform ARC and PC, and the green solid lines show their ratio.}
\label{fig:compare}
\end{figure}

Fig.~\ref{fig:compare} illustrates the comparison results for the average reflectance (top) and PDE (bottom) at a zenith angle of zero degrees and a wavelength of 420~nm in water, as a function of the AOI, for the HPK PMT (left), NNVT normal-QE PMT (middle), and NNVT high-QE PMT (right). The blue solid lines represent the results with the non-uniform PC and ARC, while the orange dashed lines correspond to the simplified model. The green solid lines in the lower panels show the ratio of the two models. Notably, the comprehensive model predicts an increase in the reflectance for the HPK PMTs, but a decrease of approximately 30\% and 40\% in the average reflectance for the NNVT normal-QE and NNVT high-QE PMTs, respectively. The comprehensive model has less than a 2\% effect on the average PDE in regions of the small AOI for all three PMT types, but a 5\% to 10\% impact on the PDE at the large AOI. It is important to note that the PDE values presented here do not include the contributions from the transmitted photons, which are event-dependent and can be accounted for through simulation.

This comprehensive PMT optical model will be integrated into the JUNO simulation program in the near future, and a more detailed analysis of its effects on energy responses will also be conducted. In addition, the outcomes derived from the comprehensive model will be compared with experimental JUNO data, and the model parameters may be further refined to improve alignment with the actual data.

\section{Conclusion}
A precise PMT optical model is essential for improving the accuracy of the energy response of the JUNO detector, thereby enhancing the sensitivity and precision of JUNO's physics outputs. In this study, we developed a comprehensive PMT optical model based on the previous research, aiming at extracting some basic optical parameters of PMTs and then precisely simulating the non-uniformity of the PDE across the PMT surface and the variations in the PDE among different PMTs. The reflectance data collected from 669 PMTs in this work as well as the PDE data gathered from both the scanning station system and the container system were utilized to constrain the thickness map of the PC and ARC for three types of 20-inch PMTs in JUNO: the HPK PMTs, NNVT normal-QE PMTs, and NNVT high-QE PMTs.

By performing a joint fit of both the reflectance and PDE data, we constructed the thickness map of the PC and ARC for each PMT. Taking into account the optical processes within the PMTs by simulation, the CE was determined as a function of the zenith angle by decomposing the PDE data. The refractive index and extinction coefficient of the PC and ARC, as well as the escape factor, were estimated over an extended wavelength range from 300 nm to 700 nm, covering the entire range of interest for JUNO.

Utilizing the new parameters for each PMT, we calculated the average reflectance and PDE, and compared these results with those from the simplified PMT optical model. The results show that the PDE undergoes changes at the level of a few percent, and the reflectance exhibits significant changes of more than 30\%. The comprehensive model will be incorporated into the JUNO simulation program, and its effects on the detector’s energy response will also be thoroughly evaluated. Furthermore, additional refinements to the model are anticipated once real JUNO data become available for comparison.

\acknowledgments
We gratefully acknowledge support from the National Natural Science Foundation of China (NSFC) under grant No. 12375196.



\bibliographystyle{JHEP}
\bibliography{biblio.bib}

\end{document}